\documentclass[sigconf]{acmart}
\usepackage{stfloats}
\usepackage{float}
\usepackage{amsmath}
\usepackage{algorithmic} 
\usepackage{array} 
\newcolumntype{P}[1]{>{\raggedright\arraybackslash}p{#1}}
\usepackage{stfloats} 
\usepackage{url} 
\usepackage{graphicx} 
\usepackage[nolist,nohyperlinks]{acronym} 
\usepackage{threeparttable} 
\usepackage{balance} 
\usepackage{multirow} 
\usepackage{xcolor}
\usepackage{flushend}
\usepackage{mathtools} 
\usepackage{wrapfig}
\usepackage{subfig}
\usepackage{array}
\usepackage{booktabs}
\usepackage{pifont}

\usepackage{amsmath} 
\usepackage{algorithmic}
\usepackage{graphicx}
\usepackage{textcomp}
\usepackage{xcolor}
\usepackage{color, soul}

\def\BibTeX{{\rm B\kern-.05em{\sc i\kern-.025em b}\kern-.08em
    T\kern-.1667em\lower.7ex\hbox{E}\kern-.125emX}}

\providecommand{\Description}[2][]{}

\usepackage{balance}

\def\BibTeX{{\rm B\kern-.05em{\sc i\kern-.025em b}\kern-.08emT\kern-.1667em\lower.7ex\hbox{E}\kern-.125emX}}

\copyrightyear{2026}
\acmYear{2026}
\setcopyright{cc}
\setcctype{by-nc-nd}
\acmConference[GeoIndustry '26]{The 5th ACM SIGSPATIAL International Workshop on Spatial Big Data and AI for Industrial Applications}{November 03--06, 2026}{Riverside, CA, USA}
\acmBooktitle{The 5th ACM SIGSPATIAL International Workshop on Spatial Big Data and AI for Industrial Applications (GeoIndustry '26), November 03--06, 2026, Riverside, CA, USA}
\acmDOI{10.1145/3849727.3856785}
\acmISBN{979-8-4007-3053-5/2026/11}

\begin{acronym}
\acro{SLAM}{Simultaneous Localization And Mapping}
\acro{BLE}{Bluetooth Low Energy}
\acro{UWB}{Ultra-Wideband}
\acro{NFC}{Near-field communication}
\acro{NLOS}{non-line-of-sight}
\acro{LOS}{line-of-sight}
\acro{EKF}{Extended Kalman Filter}
\acro{PDR}{Pedestrian Dead Reckoning}
\acro{RSS}{Received Signal Strength}
\acro{RTT}{Round Trip Time}
\acro{CNN}{convolutional neural network}
\acro{IoT}{Internet of Things}
\acro{IMU}{Inertial Measurement Unit}
\acro{RFID}{Radio Frequency Identification}
\acro{APs}{Access Points}
\acro{GNSS}{Global Navigation Satellite Systems}
\acro{LLMs}{Large Language Models}
\end{acronym}

\def \sys {\emph{TrackerNet}}

\begin{document}

\title{The Rise of Personal Trackers as the Next Infrastructure for Ubiquitous Positioning}



\author{Mohamed I. Hany}
\email{mohamed.ihany@aucegypt.edu}
\affiliation{%
  \institution{The American University in Cairo}
  \city{Cairo}
  \country{Egypt}}

\author{Hamada Rizk}
\email{hamada\_rizk@ist.osaka-u.ac.jp}
\affiliation{%
  \institution{The University of Osaka, Suita, Japan}
  \institution{RIKEN-CCS, Kobe, Japan}
  \institution{Tanta University}
  \city{Tanta}
  \country{Egypt}}

\author{Moustafa Youssef}
\email{moustafa-youssef@aucegypt.edu}
\affiliation{%
\institution{The American University in Cairo}
\city{Cairo}
\country{Egypt}}

\begin{abstract}
Reliable indoor positioning remains a central unsolved problem in spatial
computing. Existing systems depend on fixed, provider-owned
infrastructure (WiFi access points, Bluetooth beacons, UWB anchors) whose
deployment, calibration, and ownership models limit their scalability.
Meanwhile, personal trackers, wearables, smart luggage tags, and tagged
industrial assets are becoming densely embedded in the very indoor
environments where positioning infrastructure is weakest. This paper argues
that these devices point to a new class of spatial infrastructure:
\emph{non-stationary, user-owned, and semantically interpretable}. We
present \sys{}, a vision for transforming personal trackers from private
object-finders into a decentralized substrate for ubiquitous positioning.
The key shift is to stop treating anchor mobility as noise: \sys{} models
mobile anchors as probabilistic references whose stability, trust, and
semantic role are inferred over time, reframing indoor localization as
non-stationary spatial graph inference. We identify three research
frontiers (non-stationary SLAM over mobile, intermittently visible anchors;
semantic reasoning that connects ranging signals to interpretations of
anchor behavior; and consent-based, privacy-preserving governance for
cross-vendor cooperation) and a staged pathway by which the vendors and
venue operators who already run tracker networks can realize them with
academia. We do not claim a finished system; we
define a research agenda.
\end{abstract}



\begin{teaserfigure}
  \centering
  \includegraphics[width=\textwidth,height=3.3cm,keepaspectratio]{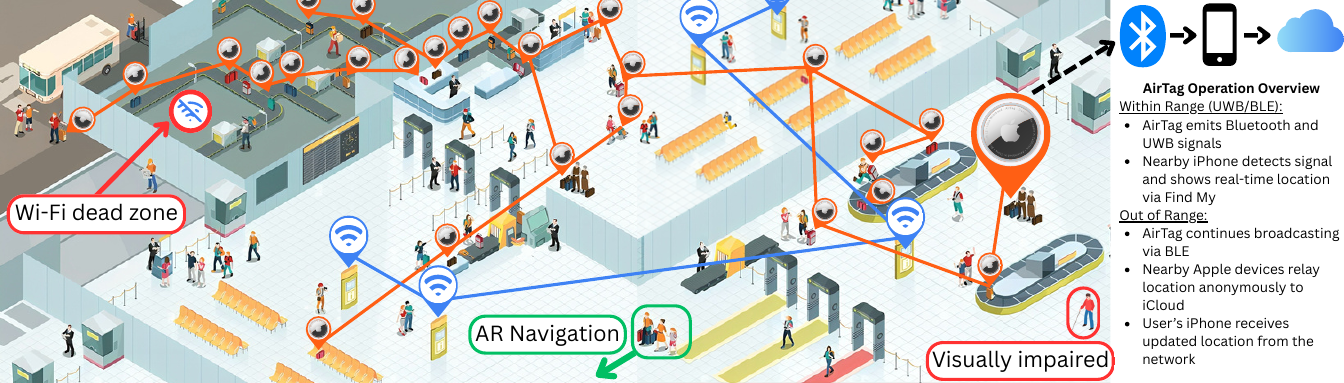}
  \Description{Isometric illustration of an airport terminal with luggage
  belts, gates, seating, and travelers. Many orange tracker markers on
  luggage, carts, and people are linked by orange lines, while a few blue
  WiFi access points are linked by blue lines. Labels mark a WiFi dead zone,
  an AR navigation user, and a visually impaired traveler. A side panel
  summarizes how an AirTag is located within and out of range.}
  \vspace{-0.2cm}
  \caption{\sys{} in a dense indoor environment. User-owned
  anchors (trackers on luggage, carts, and people) outnumber fixed access
  points; listeners observe anchors in range, a reasoning layer
  infers each anchor's stability and role, and privacy-preserving positioning
  emerges where fixed infrastructure is absent (WiFi dead zones) or
  insufficient (assistive navigation).}
  \label{fig:Vision_paper_cover}
\end{teaserfigure}


\maketitle

\section{Motivation}

Reliable indoor positioning is fundamental to human-centric spatial
computing: it underpins assistive navigation for blind and low-vision
travelers, first-responder coordination in unmapped buildings, and, at
industrial scale, asset visibility in warehouses, baggage handling in
airports, and last-meter delivery in hospitals and retail floors. Yet despite decades of progress
outdoors, enclosed spaces still rely on a fragmented patchwork of WiFi
\ac{APs}, \ac{BLE} beacons, and \ac{UWB} anchors that share a fundamental
flaw: they are static, provider-owned infrastructures, costly to deploy
per site and requiring constant recalibration as environments
drift~\cite{zafari2019survey}, while high-precision
alternatives such as dense UWB deployments~\cite{grosswindhager2019snaploc}
and 3D LiDAR sensing~\cite{yamaguchi2024towards,okochi2022fly,ukyo2023pedestrian}
achieve strong accuracy but require dedicated, costly hardware at every site.
Above all, they produce raw coordinates with no understanding of the
dynamic, human-centric spaces they measure.

A new kind of substrate is emerging that could close this gap. Personal
trackers (e.g., Apple AirTags, Samsung SmartTags)~\cite{ibrahim2023tag},
together with wearables, smart luggage tags, e-bikes, and tagged
inventory, place hundreds of millions of user-owned, radio-ranging
objects inside the very spaces fixed infrastructure struggles to
cover~\cite{hernandez2023analysis}. Industry is already
building on them: airlines accept shared tracker locations to trace
mislaid baggage~\cite{united2024airtag}, and logistics operators use
them for reverse-logistics and last-mile
visibility~\cite{noever2025airtag}. Nor are they hypothetical as
positioning anchors: the UbiLoc system shows that repurposing commercial
trackers as \emph{stationary} anchors achieves median localization error as low as 26\,cm
without manual calibration~\cite{hany2024airtags}. The deeper
opportunity lies beyond multilateration against whatever is nearby. Because these objects are \emph{semantically
meaningful} (a kiosk, a suitcase, a person), a positioning system can in
principle reason about what each anchor \emph{is} and how far to trust
it, with foundation
models~\cite{hesham2024localingua} as one
candidate mechanism for that interpretation.

Realizing this is not, however, a matter of engineering today's
products into place. An immediate and serious objection is that current
tracker networks are closed: Apple's Find My and Samsung's SmartThings
expose location only to a device's owner, and ranging against a
stranger's tag is neither permitted nor privacy-safe under existing
protocols. We take this objection seriously; it is why unlocking this substrate is
a \emph{research} agenda rather than a deployment task. The barrier is
not the hardware, which exists at global scale, but the missing
techniques and primitives needed to turn billions of mobile, privately
owned anchors into shared, trustworthy infrastructure. These are fundamentally \emph{spatial-computing} problems;
but because the substrate is owned by users and operated by vendors,
they cannot be solved by academia alone.

\emph{\textbf{Our vision}} is that future indoor positioning
infrastructure will not be installed; it will \emph{emerge} from the
mobile, privately owned, semantically interpretable objects people already
carry. We call the resulting substrate \sys{}: an architecture in which
dense personal-device networks act as probabilistic mobile anchors,
providing resilient, infrastructure-free localization. This paper
makes three contributions. First, it introduces the concept of
\emph{non-stationary, user-owned spatial infrastructure}. Second, it isolates the research gap this
creates (Section~\ref{sec:gap}) and outlines a representative agenda
around three pillars (Section~\ref{sec:vision}). Third, it lays out a
staged industrial pathway and the academia--industry collaborations each
stage requires (Section~\ref{sec:industry}).

\section{The Research Gap}\label{sec:gap}

Prior work establishes feasibility but treats trackers as
\emph{stationary} references and stops at raw coordinates~\cite{hany2024airtags};
language-model localization~\cite{hesham2024localingua}
still relies on fixed access points. The gap is not simply
that trackers are under-exploited for localization; it is that spatial systems lack a \emph{theory and infrastructure model}
for positioning over mobile, privately owned reference objects. Classical
indoor localization assumes that anchors are fixed, known, and
provider-controlled~\cite{zafari2019survey,rizk2024adaptability}. A
tracker-based substrate violates all three assumptions: anchors move,
appear and disappear opportunistically, and belong to owners who may not
wish to reveal their identity or trajectory. The closest work
relaxes only part of this: cooperative localization ranges mobile nodes
against one another~\cite{wymeersch2009cooperative} and \ac{SLAM} with moving
objects tracks moving landmarks~\cite{wang2007slammot}, but both assume
participants that cooperate by design, answer to one operator, and keep
stable identities; trackers are non-participating by default, owned by
strangers, and seen only through rotating identifiers. The result is a gap with three
facets, each of which today's work leaves open:

\vspace{-0.15cm}
\begin{itemize}
  \item \textbf{Geometric.} No positioning framework treats
    mobile anchors that are \emph{privately owned and non-cooperative} as a
    first-class substrate, modeling their motion as exploitable signal
    rather than suppressing it as noise.
  \item \textbf{Semantic.} No reasoning layer connects centimeter-level
    ranging to \emph{why} an anchor is moving and \emph{which} anchors
    to trust.
  \item \textbf{Governance.} No mechanism allows trackers across vendors
    and owners to cooperate as shared infrastructure with the owners'
    consent and without exposing any individual's identity or trajectory.
\end{itemize}
\vspace{-0.2cm}

Bridging this gap is fundamentally a spatial-computing problem, and we
frame it as one: \emph{how can spatial systems construct, query, reason
over, and govern a probabilistic positioning substrate whose reference
points are mobile, privately owned, semantically meaningful, and only
partially observable?}

\section{Vision and Research Directions}\label{sec:vision}

We envision \sys{}, an architecture that turns the dense fabric of
personal trackers into a self-healing positioning substrate. It comprises
three functional layers: \emph{(1) mobile anchors} (physical layer), the
global deployment of \ac{UWB}/\ac{BLE} trackers and related devices,
treated as transient, probabilistic reference points rather than fixed
infrastructure, alongside mains-powered devices with the same
radios, such as smart speakers and cameras, that act as naturally stable
anchors; \emph{(2) opportunistic listeners} (observation
layer), user devices such as phones or wearables that passively observe
whichever anchors happen to be in radio range at a given moment; and
\emph{(3) a reasoning layer} that interprets those observations, inferring
each anchor's stability and semantic role
(Figure~\ref{fig:Vision_paper_cover}). Unlike Find My, \sys{}
does not need anchors to know their own positions: listeners range to them
directly, anchor positions are estimated jointly with the listener's
(Eq.~\ref{eq:objective}), and a few known references, such as access points
or surveyed assets, fix the global frame. Trackers thus complement existing
infrastructure, adding density where access points are sparse and \ac{UWB}
ranging far finer than WiFi signal strength~\cite{zafari2019survey}. \emph{Self-healing} is meant precisely:
accuracy degrades locally when anchors move or leave, and recovers as new
observations arrive, without recalibration or site surveys.

\subsection{Non-Stationary SLAM: Mobility as a Feature}\label{subsec:slam}
The foundational challenge is geometric: classical \ac{SLAM} assumes
stable landmarks, but trackers ride on luggage, vehicles, and people. The conventional response is
to treat anchor motion as error and filter it out, but in a substrate
where most anchors move, this discards the very signal the system runs
on. We argue for the opposite: \emph{treating anchor mobility as an
exploitable feature}. The research problem is to develop non-stationary
\ac{SLAM} frameworks that simultaneously estimate user trajectories and
model the stochastic motion of hundreds of opportunistic anchors,
supported by selection and weighting modules that identify
``stable-enough'' anchors in real time and self-correcting algorithms
that distinguish multipath noise from genuine landmark motion.

One representative formulation estimates listener and
anchor trajectories jointly. Let $\mathbf{x}_t$ be the listener's position,
$\mathbf{a}_t^i$ that of anchor $i$, $\mathcal{A}_t$ the anchors actually
ranged at time $t$, and $s_t^i\in[0,1]$ a stability score the reasoning
layer infers from context and motion history:
\begin{equation}\label{eq:objective}
\begin{aligned}
\min_{\mathbf{x},\mathbf{a}}\;&\sum_{t}\sum_{i\in\mathcal{A}_t}
\rho\Big(\tfrac{r_t^i-\lVert\mathbf{x}_t-\mathbf{a}_t^i\rVert}{\sigma_r}\Big)
+\sum_{i,t}\frac{\lVert\mathbf{a}_t^i-\mathbf{a}_{t-1}^i\rVert^2}{\sigma^2(s_t^i)\,\Delta t_t^i}\\
&+\sum_{t}\lVert\mathbf{x}_t-f(\mathbf{x}_{t-1},\mathbf{u}_t)\rVert^2_{\Sigma_u},
\end{aligned}
\end{equation}
where $r_t^i$ is a measured range, $\rho$ a robust loss, $\sigma(\cdot)\geq\sigma_{\min}>0$
decreases with stability, $\Delta t_t^i$ is the time since anchor $i$ was
last seen, and $f$ propagates the listener from inertial input
$\mathbf{u}_t$. With all $s_t^i=1$ this approaches landmark \ac{SLAM}, the
stationary regime of~\cite{hany2024airtags}; lower scores let an anchor
drift while still constraining the listener over short windows, so
semantics, not geometry alone, decides which anchors the estimate trusts.
The pillar's open questions become properties of Eq.~\ref{eq:objective}:
how reliably and quickly $s_t^i$ can be inferred, how its errors propagate
to $\hat{\mathbf{x}}_t$, and how much accuracy is lost when identifier
rotation cuts the motion term.

While UbiLoc shows that tens-of-centimeters accuracy (26\,cm median)
is achievable with \emph{stationary} trackers~\cite{hany2024airtags},
maintaining it as the anchor graph itself moves is the open frontier of
this pillar. Energy sharpens it: coin-cell
tags built for long battery life advertise over \ac{BLE} only periodically
and range over \ac{UWB} only on demand, so a walking carrier can move an
anchor by meters between observations. Motion must therefore be modeled
rather than averaged away, and choosing $\mathcal{A}_t$ means spending
scarce \ac{UWB} exchanges on the anchors most worth ranging.

Two further questions decide whether the geometry holds in practice.
The first is \emph{density}: multilateration needs three or four
well-conditioned anchors in range, and nobody yet knows where, and when,
consumer trackers reach that density. The open problem is to characterize
anchor density as a spatiotemporal field (by venue type, time of day,
and floor), and with it the coverage gain over existing
infrastructure, and to design estimators that degrade gracefully, falling
back to inertial dead reckoning (the last term of
Eq.~\ref{eq:objective}) as density drops, rather than assuming
coverage. Industrial venues are the natural starting point precisely
because their tagged assets are dense, persistent, and owned by a single
operator. The second is \emph{verticality}: buildings are multi-level,
and ranging alone cannot distinguish an anchor one floor above from one
ten meters away on the same floor. Non-stationary \ac{SLAM} must
estimate anchor altitude jointly with position, fusing barometric and
inertial cues with the semantic layer (an anchor that just rode an
elevator changed floors), and treat floor transitions as part of the
anchor motion model rather than as outliers.

Finally, a transient cluster of listeners and anchors must \emph{agree}
on a coordinate through federated fusion; unlike established distributed
\ac{SLAM}, consensus here must be reached over an anchor set that is
itself untrusted, mobile, and uncertain.
\emph{Grand challenge: can a positioning system sustain sub-meter accuracy
when 70\% of its anchors are mobile, privately owned, and intermittently
visible?}

\subsection{Semantic Trust: From Ranging to Reasoning}
The second facet is semantic: a conventional system can measure where an
anchor is, but not \emph{what it
is} or \emph{how far to trust it}. We call the missing quantity
\emph{semantic trust}: an inferred estimate of an anchor's reliability as
a spatial reference, given what it appears to be attached to and how it
has behaved over time, a notion distinct from auditing a localization
model's outputs~\cite{abduaguye2024trust}. The first is \emph{motion-semantics classification}: inferring from
an anchor's motion profile and signal context (fusing \ac{UWB} ranging,
\ac{BLE} proximity, and inertial cues) that a tracker is fixed to a
kiosk, riding a shuttle, or carried by a pedestrian, and feeding that
inference back as $s_t^i$ in Eq.~\ref{eq:objective}. This is a
learning problem over spatiotemporal signals, likely solved by compact
motion models rather than language models; reliable inference of motion
semantics from noisy signals is today unvalidated and is among the central
risks of this agenda. The object of classification is the \emph{anchor's role as a
reference} (fixed, vehicular, pedestrian), never the identity or
destination of a person carrying it. The second problem is the
\emph{interface}: translating the resulting spatial state into guidance
humans can act on. This is where language and spatial foundation models
clearly earn their place: not as the positioning engine, but as the layer
that turns a trust-weighted spatial graph into actionable interaction.

Scale introduces a hard systems constraint: in dense settings,
thousands of \ac{BLE}/\ac{UWB} broadcasts arrive simultaneously, so the
reasoning layer must prioritize the signals relevant to the user's current
goal, on-device and within interactive latency. The most demanding
application is a ``spatial narrator'' for blind and low-vision users that
detects and announces subtle changes such as a moved chair or a temporary
queue.
\emph{Grand challenge: can an anchor's semantic role and stability be
inferred from motion and signal context with high enough precision and
recall, and generalize across unseen venues and device types, to be
trusted as a constraint on positioning?}

\subsection{Responsible Spatial Infrastructure}\label{subsec:gov}
The third facet is governance, the precondition for the other two:
infrastructure built from strangers' devices must be trusted by their
owners as much as by its users. We treat four problems as \emph{design constraints}.

The first is \emph{consent and legal standing}. An anchor participates
only if its owner has explicitly opted in; non-participation is the
default, opt-in is revocable, and a non-participating anchor is as
invisible to the substrate as it is to strangers today. Whether such
consent can be given lawfully is itself open: a rotating identifier can
still be personal data under regimes such as the GDPR, so the purpose for
which an anchor's radio presence is used must be bounded and auditable.
In the operator-owned settings of Section~\ref{sec:industry}, consent
reduces to one organizational decision, a reason to start there.

The second is the \emph{privacy--utility tension}: a listener must be
able to range against a consenting stranger's tracker without ever
learning or storing that owner's persistent identity or trajectory.
Today's tracker networks rely on encrypted, rotating identifiers, but
using them as positioning anchors reintroduces exactly the persistent
tracking risk those rotations were designed to
prevent~\cite{shafqat2023track}. We see a research frontier in
\emph{zero-knowledge spatial proofs}: secure multi-party computation,
private set intersection, and distance-bounding protocols already let
distrustful parties establish proximity without disclosing private inputs. The open problem is to compute a
multilateration solution over committed, rotating anchor identifiers such
that the listener learns its own position but nothing about the
contributing anchors (in Eq.~\ref{eq:objective}, revealing
$\hat{\mathbf{x}}$ while keeping $\hat{\mathbf{a}}$ hidden), within
real-time latency. We do not claim this is
solved; we argue it is plausible, and a prerequisite.

The third is \emph{dual use and misuse resistance}. Any system that
localizes via strangers' devices could be turned toward surveillance,
and the risk is asymmetric: a venue operator or state can deploy far
more listeners than an individual can. We hold that \sys{} is acceptable only
if it makes covert tracking of a person \emph{harder} than today's
networks do: the protocol must be one-directional (a listener learns
where \emph{it} is, never where an anchor has been), observations must be
unlinkable across identifier rotations, and the right to deploy listeners
at scale must itself be governed. Whether these properties can be enforced
technically rather than promised is open; if they cannot, it
should not be built.

The fourth is \emph{uncooperative and adversarial anchors}. An
uncooperative anchor is not a fault but the default, handled by the
density-aware estimation of Section~\ref{subsec:slam}. A malicious
anchor, one that spoofs its ranging or motion signature to corrupt a
fix, is a Byzantine problem: robust estimators must bound any single
anchor's influence ($\rho$ in Eq.~\ref{eq:objective}), cross-check its reports against independent
listeners, and, via privacy-preserving reputation, earn trust for the
\emph{spatial quality} of a contribution (stability, coverage, ranging
consistency) without exposing the owner's history. The same mechanism
serves \emph{incentive alignment}, rewarding honest, stable anchors as
density grows. Finally, \emph{cross-vendor interoperability}: the 2024
Apple--Google specification for detecting unwanted trackers shows rival
vendors can agree on cross-platform tracker protocols; a spatial commons
needs an equivalent for spatial metadata, and a stewardship model that
keeps it open and resilient to monopolistic control.
\emph{Grand challenge: can a consenting stranger's anchor contribute to a
position fix with provably zero leakage of its owner's identity or
trajectory, at interactive latency?}

\section{Industrial Impact and Pathway}\label{sec:industry}

Unusually for a positioning vision, \sys{}'s substrate is already
deployed and owned: a handful of vendors operate hundreds of
millions of trackers, and tagged assets concentrate in exactly the
venues (warehouses, airports, hospitals, distribution centers) where
operators pay most for fixed infrastructure. The industrial value is
direct: positioning with no site survey or per-site CapEx that
self-heals as layouts change. A
staged pathway lets industry realize value early while the harder
research matures.

\emph{Stage 1: Operator-owned anchors, single vendor.} A warehouse or
airport already tags its own pallets, carts, and equipment; using them as
mobile anchors for the operator's own devices raises no consent or
cross-vendor barrier and can start today: UbiLoc's
calibration-free 26\,cm result sets the stationary baseline, and this is
where the mobile case is first measured. This stage tests
non-stationary \ac{SLAM}, density, and multi-floor estimation, and its
natural output is a \emph{benchmark dataset} of
anonymized ranging and motion traces from live industrial venues, data that
academia cannot collect alone.
\emph{Stage 2: Opt-in, cross-vendor, within a venue.} Employee and
customer devices join under explicit, revocable consent, exercising the
mechanisms of Section~\ref{subsec:gov} at bounded scale under one legal
entity. \emph{Stage 3: A spatial commons.}
Vendor-agnostic standards and stewardship extend the substrate across
venues and owners.

Academia contributes the estimation
theory, trust-inference models, and cryptographic primitives that make a
deployment lawful and robust; industry contributes devices, venues,
anonymized data, and a seat at the standards table. For vendors the
payoff is a new service tier on existing hardware; for operators,
positioning without infrastructure; for researchers, a substrate for
studying non-stationary spatial inference at real scale.

\section{Conclusion}\label{sec:conclusion}
We have argued that personal trackers point to a non-stationary,
user-owned, semantically interpretable class of spatial infrastructure,
reached through the industry that already deploys it: positioning
infrastructure that is not installed, but carried.


\bibliographystyle{ACM-Reference-Format}
\bibliography{references} 

\end{document}